PD1

# High-efficiency single photon detector combined with an ultra-small APD module and a self-training discriminator for high-speed quantum cryptosystems

Seigo Takahashi (1), Akio Tajima (1), Akihisa Tomita (2,3)
(1) System Platforms Res. Labs. NEC Corporation
1753 Shimonumabe Nakahara-ku Kawasaki Kanagawa, Japan
(2) Nano Electronics Res. Labs. NEC Corp., (3) ERATO-SORST, JST

*Abstract:*
*A single-photon avalanche detector (SPAD) for high-speed quantum-key generation has successfully been developed. It has the highest photon detection repetition frequency and the lowest dark count rate in the world, as a board-mountable sub-system. The SPAD consists of an ultra-small dual-avalanche photodiode (APD) module and a novel discriminator. The APD module design is consistent with cooling capability and high-frequency characteristics. The new module has a 3 GHz bandwidth enabling 1 GHz gate-pulse repetition. The bandwidth is extended 15-fold relative to the most wideband peltier cooled APD module. The discriminator has a self-training mechanism to compensate charge pulse. Dark count rare of the SPAD is reduced $1/10^{th}$ relative to the lowest dark count single photon detector. The SPAD allows 3.2-fold multiplying the quantum key generation rate in theoretical estimation.*
© 2007 Microoptics Group (OSJ/JSAP)

## 1. Introduction and issues

Various security attacks threaten the security of the Internet. Quantum cryptography is an attractive candidate for keeping data confidential because it exploitation of quantum theory guarantees perfect security. Actual quantum cryptosystems with high key generation rates and long-distance transmission capabilities have been reported [1-3]. A single photon avalanche detector (SPAD) is the most important component to be used as the key generation component. The system requires small size, low power consumption, and high key-generation rate are requirements for the SPAD consisting of APD modules and a discriminator.

In particular, the APD module must be small, have low power consumption and a cooling capability, and must be simultaneously capable of long-time continuous operation and high repetition frequency operation. The discriminator has a problem of detection efficiency caused by differences between individual devices.

The APD module reported here consists of InGaAs-APDs which have sensitivity at 1.55 μm and can be used for long-distance fiber transmission. It must be operated at very low temperature and in gate mode [4] to suppress the dark count rate. The APD module should be cooled to around -70°C and have several GHz bandwidth in order to a high gate repetition frequency. Moreover, the APD module should be about the same size as a butterfly-type module for telecom equipment. A butterfly-type module with built-in peltier cooler is developed [5]. However, it has not enough bandwidth. To overcome the above problem, we developed an ultra-small APD module with a 3-GHz bandwidth and 200 K cooling capability.

The gate pulse operation induces charge pulses which degrade discrimination sensitivity. Several charge pulse compensators were reported to reduce this degradation [6-10]. However, high-frequency operation increases the charge pulse compensation error caused by differences between the individual SPAD devices (e.g. the APD chip and module). These devices have errors in their frequency response caused by manufacturing fluctuations. Precise adjustment of the discriminator to compensate for these individual differences becomes increasingly difficult as gate repetition frequency increases. To overcome the above problem, we developed a novel discriminator with a self-training charge pulse compensator.

## 2. Ultra-small APD module with high frequency bandwidth

The peltier device's cooling capability deteriorates in proportion to its thermal load. A larger thermal load requires a larger cooler, which in turn requires a larger power consumption and heat sink. Thus, the thermal load should be made as low as practical. It is better to reduce number of wirings for the electronic signal because they are the dominant thermal flux paths. On the other hand, thick wiring is required for a wideband frequency response. Therefore, the issue is designing the wires in the module with arbitration of these inconsistent requirements.

The appearance of the APD module prototype is shown in Figure 1. The body is 23.3 x 21.7 x 15 $mm^3$.

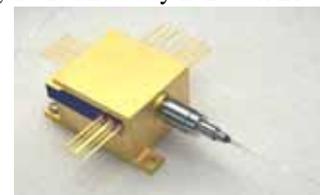

Figure 1. Appearance of the APD module

The module has the following features:
(1) Two APD chips are cooled by one peltier cooler. The

cryptosystem uses two or more APDs to generate quantum keys. This scheme can reduce total size and cooling power dissipation in the system.

(2) The module has a lens for two APD chips and two fibers. The optical beams launched from the fibers are focused on the respective APDs (Figure 2). Responsibility at -70°C was evaluated to be 1.02 ~ 1.05 A/W for both APDs operating with very little crosstalk.

(3) The module has incoming and outgoing coplanar lines as the gate pulse transmission lines. The gate pulse waveform has a bandwidth of a few GHz. Therefore, the transmission line has to terminate outside of the module to avoid putting a thermal load on the peltier cooler.

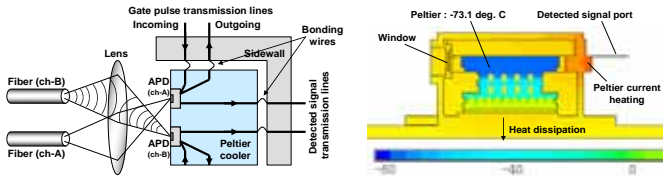

Figure 2. Internal structure of the APD module

Figure 3. Thermal simulation result

2.1 Thermal design

The total thermal flux to the cooled area was evaluated with a finite element method thermal simulator. The maximum allowable thermal flux was calculated to be up to 250 mW for APD chips cooled to -70 deg. C (Figure 3). The total thermal flux through the electronic wiring is around 200 mW, which is small enough to evacuate with the peltier cooler.

An experiment confirmed that the prototype modules could be cooled to -73 deg. C (200 K).

2.2 RF design

The module's bandwidth must be about 3 GHz because it is necessary to apply a gate pulse with a 1-GHz repetition rate to the APD chip. The gate pulse transmission line has two bonding wires (Figure 3). The bandwidth is affected by multiple reflections between the wires. An FDTD simulation indicated that a bonding wire length of less than 5 mm would ensure a 3 GHz bandwidth.

An experiment confirmed that the prototype module had a bandwidth of up to 3 GHz. The bandwidth allows the gate pulse repetition frequency to increase 15-fold in comparison with the previous most wideband peltier cooled APD modules [11].

## 3. Discriminator with self-training charge pulse compensator

As mentioned above, the discriminator's sensitivity is affected by the differences between individual devices. Figure 4 shows a novel discriminator with a self-training charge pulse compensation scheme. Digital signal processing is applied to the compensator, which consists of an A/D converter (ADC) and logic circuit. The charge pulse is sampled by the ADC. The average charge pulse waveform is digitally obtained from past several cycles of the charge pulse without precise circuit adjustment. The most recent sampled data are discriminated by using the averaged charge pulse. This scheme is not affected by individual device differences because it uses its own waveforms.

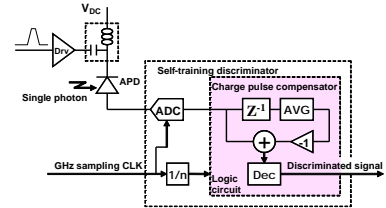

Figure 4. Architecture of novel discriminator with self-training charge-pulse compensator

## 4. Photon detection experiments

The photon detection characteristics of the APD module prototype were evaluated with the discriminator. Photon detection efficiency ($P_{PD}$) and dark count probability ($P_{DK}$) were measured versus discrimination level ($V_{th}$) (Figure 5). The $P_{DK}$ was about $1/10^{th}$ lower than that of the previous most low dark count SPAD at the same $P_{PD}$ [6]. The $P_{DK}$ reduction allows 3.2-fold increasing the quantum key generation rate in theoretical estimation [12] on our system [1].

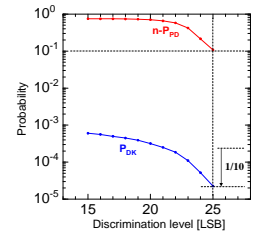

Figure 5. Single photon detection characteristics

## 5. Conclusions

A board-mountable single photon avalanche detector (SPAD) has been developed using a small APD module and a novel discriminator. The module has a 3 GHz gate pulse bandwidth and can be cooled to 200 K. A SPAD combining an APD modules and a self-training discriminator allows 3.2-fold multiplying the quantum key generation rate.

**Acknowledgements**
Part of this work was supported by the National Institute of Information and Communication Technology (NICT).